\documentclass[a4paper,floatfix,11pt]{article}
\usepackage{graphicx}
\usepackage{amsmath}
\usepackage{dcolumn}
\usepackage{bm}
\usepackage{lineno}
\usepackage[english,spanish,activeacute]{babel}

\def \gcm {$\mathrm{\ g\ cm^{-2}}$}

\def \nutau {$\nu_\tau~$}
\def \vavrg {$\langle V \rangle $}
\def\Offline{\mbox{$\overline{\textrm{Off}}$\hspace{.05em}\raisebox{.4ex}{$\underline{\textrm{line}}$}} }

\def \pao {Pierre Auger Observatory}

\newcommand{\degree}{\ensuremath{^\circ}}
\title{\Huge Ultrahigh Energy Neutrinos \\ 
at the Pierre Auger Observatory}

\author{The Pierre Auger Collaboration
\footnote{Correspondence to
The Pierre Auger Collaboration; auger spokesperson@fnal.gov}}

\begin{document}

\selectlanguage{english}
\maketitle

\par\noindent
{\bf The Pierre Auger Collaboration} \\
P.~Abreu$^{65}$, 
M.~Aglietta$^{53}$, 
M.~Ahlers$^{95}$, 
E.J.~Ahn$^{83}$, 
I.F.M.~Albuquerque$^{16}$, 
D.~Allard$^{30}$, 
I.~Allekotte$^{1}$, 
J.~Allen$^{87}$, 
P.~Allison$^{89}$, 
A.~Almela$^{11,\: 7}$
% ...
\\

\par
\par\noindent
$^{1}$ Centro At\'{o}mico Bariloche and Instituto Balseiro (CNEA-UNCuyo-CONICET), San Carlos de Bariloche, Argentina \\
% ...
$^{7}$ Instituto de Tecnolog\'{\i}as en Detecci\'{o}n y Astropart\'{\i}culas (CNEA, CONICET, UNSAM), Buenos Aires, Argentina \\
% ...
$^{11}$ Universidad Tecnol\'{o}gica Nacional - Facultad Regional Buenos Aires, Buenos Aires, Argentina \\
% ...
$^{16}$ Universidade de S\~{a}o Paulo, Instituto de F\'{\i}sica, S\~{a}o Paulo, SP, Brazil \\
% ...
$^{30}$ Laboratoire AstroParticule et Cosmologie (APC), Universit\'{e} Paris 7, CNRS-IN2P3, Paris, France  \\
% ...
$^{53}$ Istituto di Fisica dello Spazio Interplanetario (INAF), Universit\`{a} di Torino and Sezione INFN, Torino, Italy \\
% ...
$^{65}$ LIP and Instituto Superior T\'{e}cnico, Technical  University of Lisbon, Portugal \\
% ...
$^{83}$ Fermilab, Batavia, IL, USA \\
% ...
$^{87}$ New York University, New York, NY, USA \\
% ...
$^{89}$ Ohio State University, Columbus, OH, USA \\
% ...
$^{95}$ University of Wisconsin, Madison, WI, USA \\
% ...
\par\noindent
(\ddag) Deceased \\
(a) at Konan University, Kobe, Japan \\
(f) now at University of Maryland \\
(h) now at NYU Abu Dhabi \\
(i) now at Universit\'{e} de Lausanne \\

\begin{abstract}
The observation of ultrahigh energy neutrinos (UHE$\nu$s) has become a priority in experimental astroparticle
physics. UHE$\nu$s can be detected with a variety of techniques. In particular, neutrinos can interact in the 
atmosphere (downward-going $\nu$) or in the Earth crust (Earth-skimming $\nu$), producing air showers that can be observed
with arrays of detectors at the ground. With the Surface Detector Array of the Pierre 
Auger Observatory we can detect these types of cascades. 
The distinguishing signature for neutrino events is the presence of very inclined showers produced 
close to the ground (i.e. after having traversed a large amount of atmosphere).
In this work we review the procedure and criteria established to search for 
UHE$\nu$s in the data collected with the ground array of the Pierre Auger Observatory. 
This includes Earth-skimming as well as downward-going neutrinos.
No neutrino candidates have been found, which allows us to place competitive limits 
to the diffuse flux of UHE$\nu$s in the EeV range and above.  
\end{abstract}

\section{Introduction}
\label{sec:intro}

The observation of ultrahigh energy cosmic rays (UHECR) of energy 1~$-$ 100~EeV ($10^{18}-10^{20}$ eV) has
stimulated much experimental as well as theoretical activity in the field of Astroparticle Physics 
\cite{Nagano_Watson,Kotera-Olinto}. 
Although many mysteries remain to be solved, such as the origin of the UHECRs, their 
production mechanism and composition,  
we know that it is very difficult to produce these energetic particles without associated fluxes 
of ultrahigh energy neutrinos (UHE$\nu$s) \cite{Halzen-Hooper}. 

In the so-called ``bottom-up" models, protons and nuclei are accelerated in astrophysical 
shocks, where pions are believed to be produced by cosmic ray interactions with matter or radiation at the source \cite{Becker}. 
In the so-called ``top-down" scenarios, protons and neutrons are produced from quark and gluon fragmentation, a
mechanism which is known to produce much more pions than nucleons \cite{Bhatta}. Furthermore,
protons and nuclei also produce pions in their unavoidable interactions responsible for the Greisen-Zatsepin-Kuzmin (GZK)
cut-off \cite{GZK,BZ}. 
The flux of UHECRs above $\sim 5\times 10^{19}$ eV is known to be largely suppressed with respect to that at lower energies, 
a feature seen in the UHECR spectrum \cite{HiRes_spectrum,Auger_spectrum} that is compatible with the interaction of UHECRs  
with the cosmic microwave background (CMB) radiation.
If the primaries are protons, the interaction responsible for the GZK effect is photopion production, 
and the decays of the charged pions produce UHE neutrinos. 
However, their fluxes are uncertain \cite{Becker}, and if the primaries are heavier nuclei, 
the UHE$\nu$ yield would be strongly suppressed \cite{Kotera_GZK}.

The observation of UHE neutrinos could provide important hints to the origin of UHECRs \cite{Stanev,Allard}. 
Unlike cosmic rays, neutrinos point directly to the source where they were produced, without being deflected
by Galactic and extragalactic magnetic fields. Unlike photons they 
travel undisturbed from the sources carrying a ‘footprint’ of the production model.

High energy neutrinos can be detected with a variety of techniques \cite{Veronique,Anchordoqui-Montaruli}.
In particular they can be observed with arrays of detectors at ground level that are currently 
being used to measure extensive showers produced by cosmic rays \cite{zas_tau}.
The main challenge in this technique lies in separating showers
initiated by neutrinos from those induced by regular cosmic rays. It was suggested in the
1970s that this could be done at high zenith angles \cite{Berezinsky_HAS} because the 
atmosphere slant depth provides a quite large target for neutrino interactions. 
The idea is that neutrinos, having very small cross-sections, can interact at any point along their
trajectories, while protons, nuclei or photons interact shortly after entering the atmosphere.  
The signature for neutrino events is thus inclined showers that interact deep in the atmosphere. 

Inclined showers were first observed in the 1960s by several groups \cite{AKENO_HAS,Haverah_Park_HAS}. 
With the Surface Detector Array (SD) of the Pierre Auger Observatory~\cite{EApaper} 
we can detect inclined showers and identify neutrinos with energies typically above 0.1 EeV. 
There are two ways of performing this task: 
\begin{enumerate}

\item
Neutrinos of all flavours can
collide with nuclei in the atmosphere and induce an extensive air shower 
close to the ground~\cite{nu_down_theory_Auger}. In this so-called ``downward-going"
neutrino channel, both charged current (CC) and neutral-current (NC)
interactions contribute to the neutrino event rate.

\item
Neutrinos of tau flavour ($\nu_\tau$) are expected to be most sensitively 
observed through the detection of showers induced by the decay products of an emerging
$\tau$ lepton, after the propagation and interaction of an upward-going $\nu_{\tau}$ 
inside the Earth \cite{Bertou_nutau,Fargion_nutau}. 
This ``Earth-skimming" channel benefits from the long range of the $\tau$ lepton
($\sim$ 10 km for the shower energies relevant in this analysis)
which sets the scale of the effective volume. 
Only charged-current interactions of $\nu_{\tau}$ are relevant in this case. 

\end{enumerate}

In both the Earth-skimming and downward-going channels the
showers can be identified and separated from cosmic ray induced showers 
with the SD of the Pierre Auger Observatory if the zenith angle is large enough, typically
larger than $\sim 65^\circ$--$75^\circ$. A number of properties of the shower front,
 mostly stemming from the time distribution of the shower
particles, can be used to distinguish neutrino-induced showers. 
As shown in Sec. \ref{sec:nu_selection},
even though the criteria to identify neutrinos in both channels
being based on similar ideas and variables, two different analyses were designed.
The main reason for that concerns background reduction.
The Earth-skimming neutrino search is restricted to a very narrow
angular range where the background of nucleonic showers is expected
to be very small. On the other hand in the broader angular range of the downward-going neutrino search 
the background contamination is expected to be larger, and the selection
criteria need to be more restrictive. This calls for specific algorithms and methods, 
capable of optimizing the separation of neutrino-induced showers from nucleonic ones 
as will be explained later in the paper. 

In this work we review the procedure to search for UHE$\nu$s with the SD of the Auger Observatory,
for both the Earth-skimming and downward-going channels. In Sec. \ref{sec:PAO} we give a brief overview
of the SD of the Pierre Auger Observatory. In Sec. \ref{sec:general} we concentrate on the general strategy
to search for UHE$\nu$s. Sec. \ref{sec:sims} is devoted to describe the simulations of neutrino-induced
showers crucial to establish selection criteria and to compute the exposure to UHE$\nu$s
which is reported in Sec. \ref{sec:exposure}. In Sec. \ref{sec:nu_selection} we give a detailed 
description of the neutrino selection criteria. When these criteria are applied blindly to the data collected
at the SD no candidates are found.
The resulting limits to the diffuse flux of UHE$\nu$s are presented in Sec. \ref{sec:results}.
Finally, in Sec. \ref{sec:conclusions} 
we summarize the paper and give some prospects for future observations.

\section{The Pierre Auger Observatory}
\label{sec:PAO}

The Pierre Auger Observatory \cite{EApaper} is a hybrid UHECR 
detector combining an array of particle detectors 
at ground level, and 24 fluorescence
telescopes housed in four buildings, for redundancy and calibration.
It is located near the town of Malarg\"ue, in 
the province of Mendoza in Argentina. 
In this review we focus on the surface detector array \cite{EApaper,Auger_SD} 
which is briefly described in the following. 

\subsection{The Surface Detector Array}

\begin{figure}
\begin{center}
\includegraphics[width=0.95\textwidth]{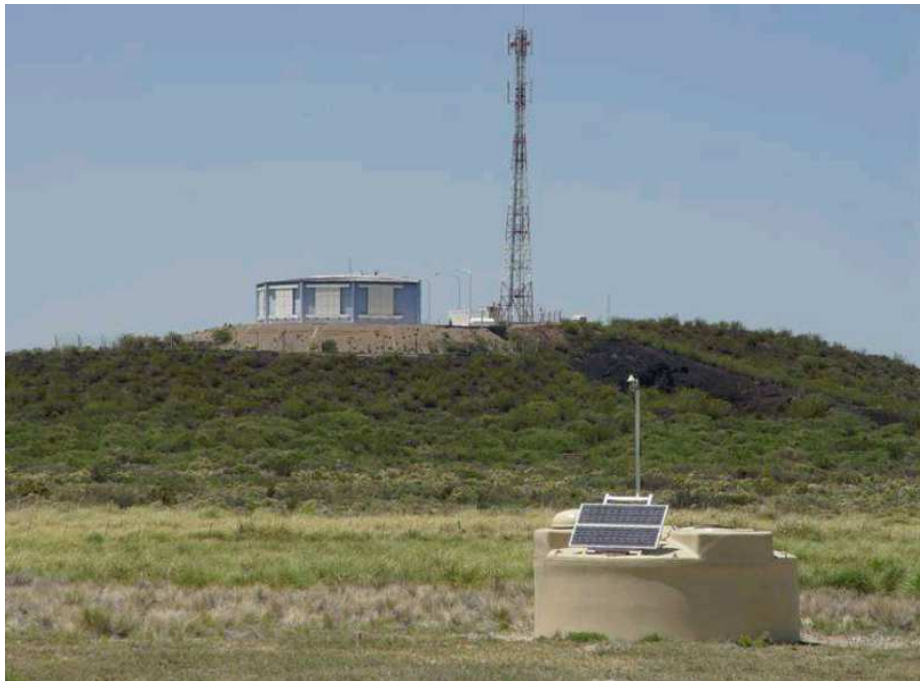} \\
\includegraphics[width=0.95\textwidth]{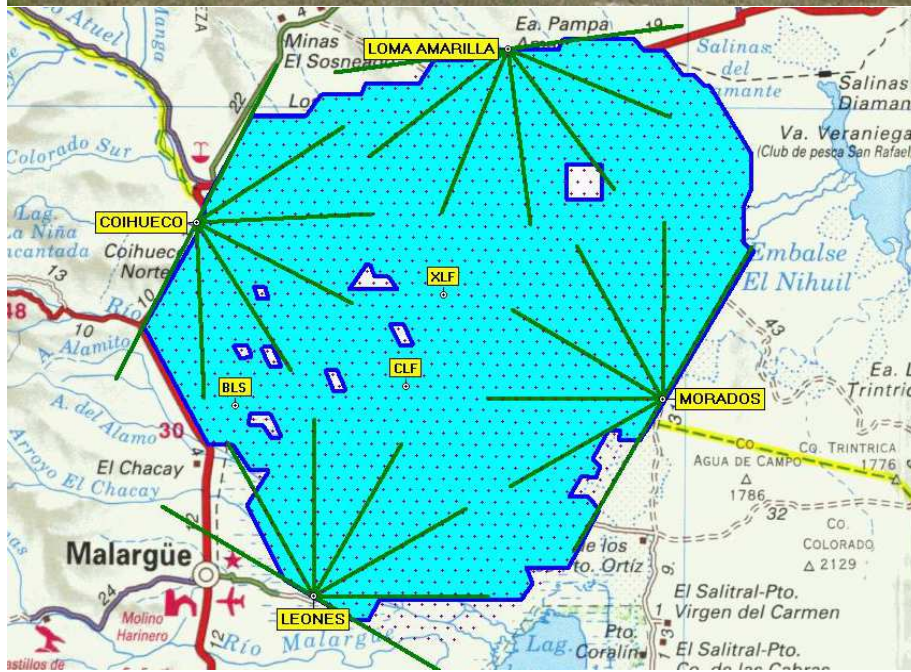}
\caption{Top panel: one of the $\sim 1600$ water Cherenkov 
stations that constitute the surface detector array of the 
\pao$~$ (forefront), and one of the four fluorescence buildings housing six
of the 24 fluorescence telescopes (background).
Bottom panel: Layout of the SD array with $\sim 1600$ water Cherenkov stations (depicted as dots),
spread over a surface of $\sim 3000~{\rm km^2}$ (blue area), with a distance between stations of 1.5 km.
The four fluorescence buildings at the edges of the Observatory are also indicated.}
\label{fig:tank}
\end{center}
\end{figure}

The surface detector array \cite{Auger_SD} consists of water Cherenkov detectors
in the form of cylinders of 3.6 m diameter and 
1.2 m height, each containing 12 tonnes of purified water.
Charged particles entering the station emit Cherenkov 
light which is reflected at the walls by a diffusive Tyvek liner,
and collected by three 9-inch photomultiplier tubes (PMT) at 
the top surface and in optical contact with the water. 
 The PMT signals are sampled by flash analog to digital 
converters (FADC) with a time resolution of 25 ns.
Each station is regularly monitored and calibrated in units of vertical
equivalent muons (VEM) corresponding to the signal produced
by a muon traversing the tank vertically through its center \cite{Auger_calibration}.
In Fig.~\ref{fig:tank} we show a picture of one of the water Cherenkov
stations. The stations are autonomous, with all their components
(PMTs, local processor, GPS receiver and radio system) 
powered by batteries coupled to solar panels. 
Once installed, the local stations work continuously
without external intervention.

The SD was completed in 2008.
There are $\sim$ 1600 water stations arranged in a triangular grid 
with 1.5 km spacing between them, spanning an almost flat surface
of $\sim 3000~{\rm km^2}$, at an approximate
altitude of 1400 m above sea level, or equivalently an atmospheric
depth $X_{\rm ground}=880$ \gcm. The layout of the SD array is
sketched in the bottom panel of Fig.~\ref{fig:tank}. 

\subsection{Surface Detector trigger}

The stations transmit information by conventional radio links
to the Central Data Acquisition System (CDAS) located in Malarg\"ue. 
There are two types of trigger conditions. A local trigger at the level 
of an individual station (second order or T2 trigger), and a global
trigger (third order or T3 trigger). The T2 trigger condition is the 
logical OR of two conditions: either a given threshold signal (3.2 VEM) is
passed in at least one time bin of the FADC trace --the so-called
``Threshold trigger"--, or a somewhat lower threshold (0.2 VEM) 
is passed in at least 13 bins within a 3 $\mu$s time window 
(i.e. 120 bins) --the so-called ``Time-over-Threshold (ToT) trigger". 
The ToT condition was designed to trigger on signals broad
in time, characteristic of the early stages of the development 
of an extensive air shower, and is crucial for neutrino identification
as explained below. The data acquisition system receives the
local T2 triggers and builds a global T3 trigger requiring a
relatively compact configuration of at least three local stations
compatible in time, each satisfying the ToT trigger, or four triggered stations 
with any type of T2 trigger \cite{Auger_trigger}.
With the completed array, the global T3 trigger
rate is about two events per minute, one third being
actual shower events at energies above $3 \times 10^{17}$ eV.

\section{Generalities of UHE Neutrino search} 
\label{sec:general}

With the SD of the Pierre Auger Observatory we can detect and identify
UHE neutrinos in the EeV range and above \cite{PRL_nu_tau,nu_tau_long,nu_down}. 
The main challenge from the experimental point of view is to identify neutrino-induced
showers in the large background of showers initiated by nucleonic cosmic rays.
The concept for identification is relatively simple.
While protons, heavier nuclei and even photons interact shortly
after entering the atmosphere, neutrinos can generate showers initiated deeply
into the atmosphere.
When considering vertical showers, even the ones initiated by protons or heavy nuclei
have a considerable amount of electromagnetic component
at the ground (``young'' shower front). 
However, when looking at high zenith angles ($\theta>75\degree$) the atmosphere is thick enough 
(thicker than about three vertical atmospheres)
so that the cosmic rays interacting high in the atmosphere have shower fronts 
dominated by muons at ground (``old'' shower front).
A neutrino with $\theta>75\degree$ interacting deep will present a young shower front and, consequently, can be distinguished.

\begin{figure*}[htb]
\begin{center}
\noindent
\begin{tabular}{cc}
\hspace{-0.5cm} \includegraphics [width=0.55\textwidth]{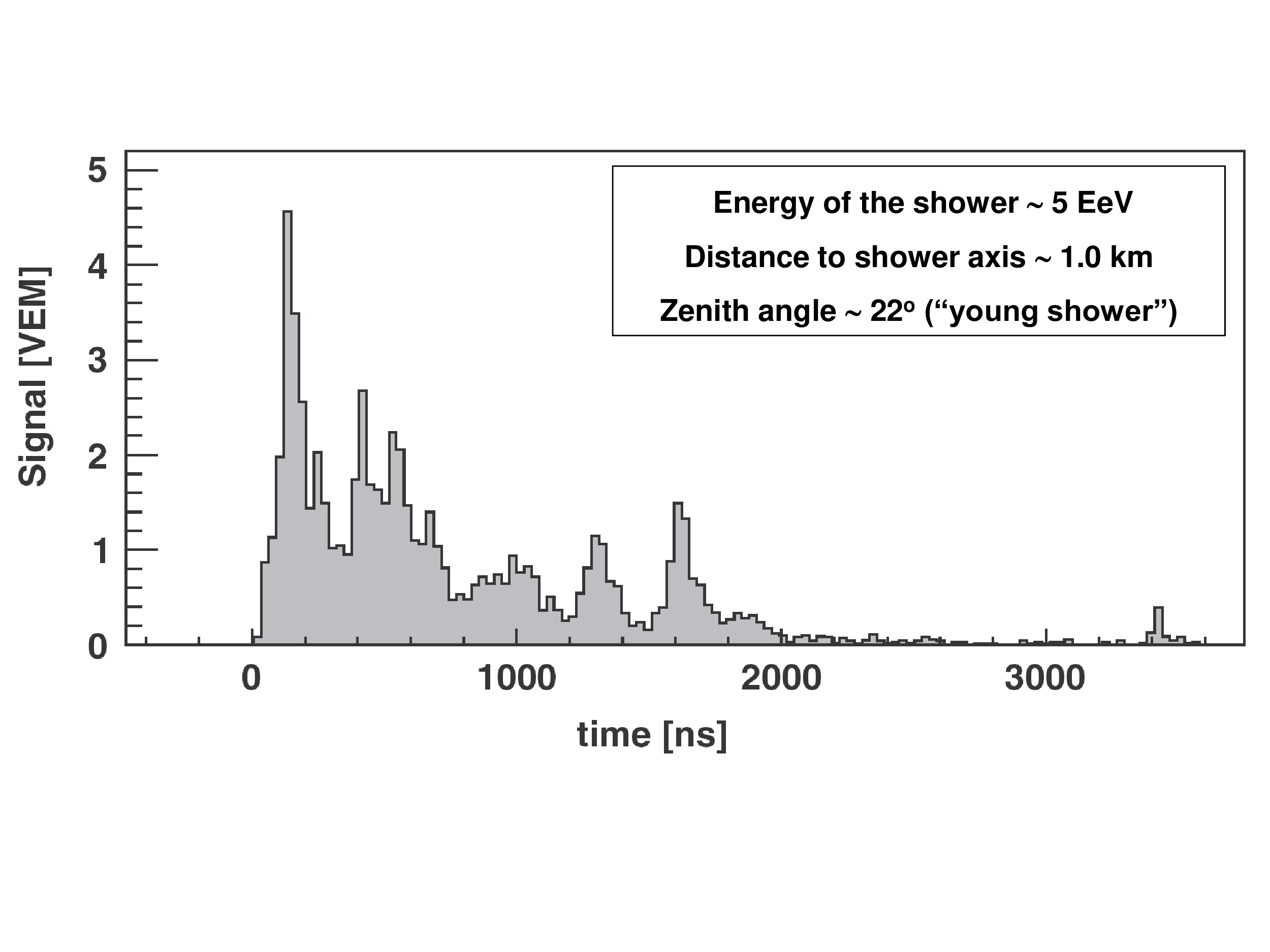} &
\includegraphics [width=0.55\textwidth]{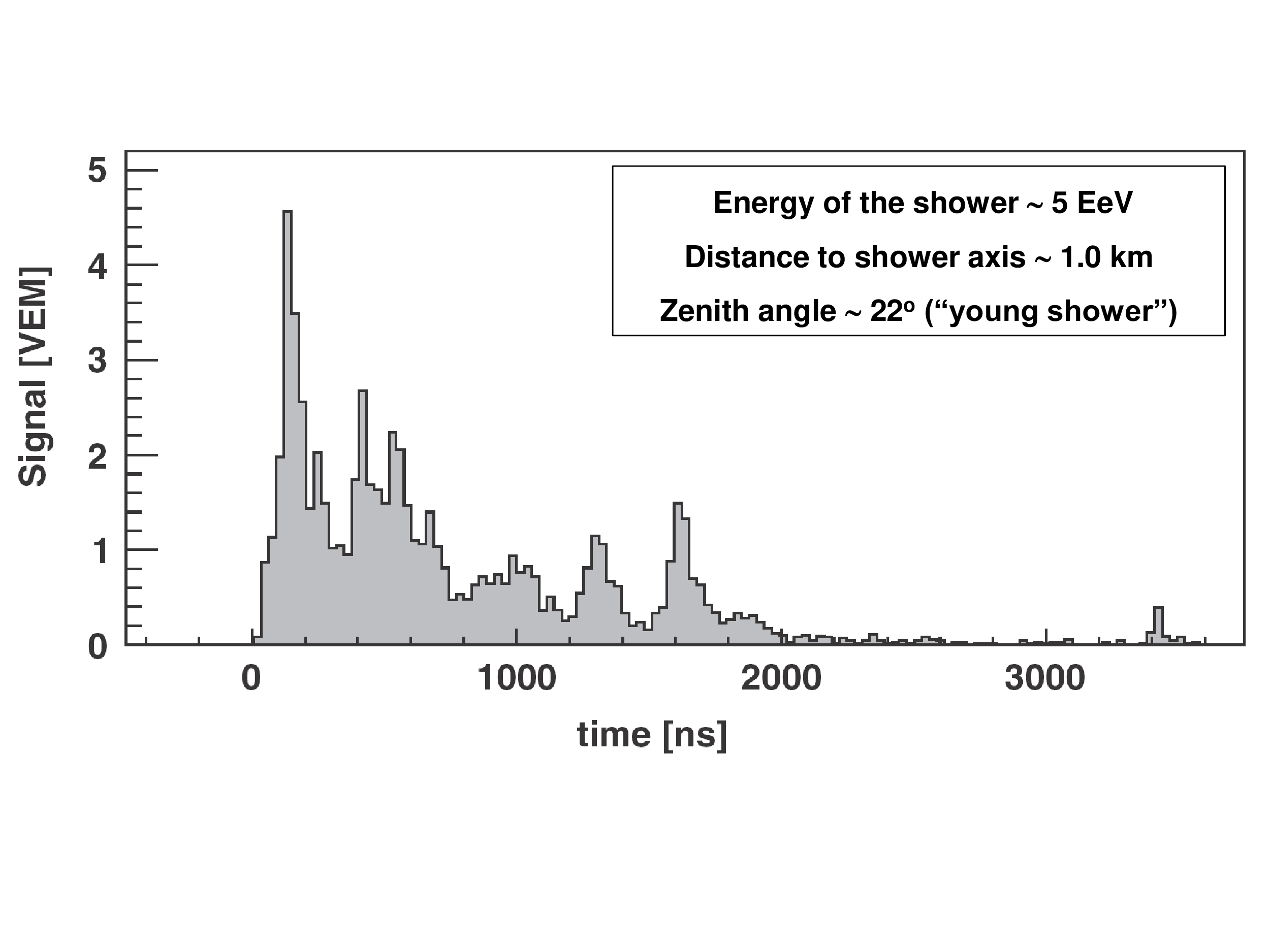}
\end{tabular}
\end{center}
\vskip -0.7cm
\caption{
FADC traces of stations at 1 km from the shower core for two real showers of 5 EeV.
Left panel: shower arriving in the early stages of development (``young'' shower).
Right panel: ``old'' extensive air shower ($\theta \sim 80^\circ$).}
\label{fig:traces}
\end{figure*}

At the SD level, young showers induce signals 
spread in time over hundreds of nano-seconds in a fraction of the stations triggered
by the shower, while old showers induce narrow signals spreading over typically
tens of nano-seconds in practically all the stations of the event.  
With the 25 ns time resolution of the FADC of the water Cherenkov stations, the 
distinction between traces induced by young and old shower fronts can be easily accomplished. 
In Fig.~\ref{fig:traces} we show an example of those two types of traces. 

With this simple idea, we can search for two types of neutrino-induced showers
at the surface detector array of the Pierre Auger Observatory, namely:

\begin{enumerate}

\item Earth-skimming showers induced by tau neutrinos (\nutau)
that travel in the upward direction with respect to the vertical to ground.
\nutau can skim the Earth's crust and interact relatively close to the surface 
inducing a tau lepton which escapes the Earth and decays in flight in the atmosphere,
close to the SD. 

Typically, only Earth-skimming \nutau-induced showers with zenith 
angles $90^\circ < \theta < 95^\circ$ may be identified. 

\item Showers initiated by any neutrino flavour moving down at large angles with respect
to the vertical at ground that interact in the atmosphere close 
to the surface detector array. We include here showers induced
by \nutau interacting in the mountains surrounding the \pao. Although  
this latter process is exactly equivalent to the
``Earth-skimming'' mechanism, it is included in this class because such showers 
are also going downwards. In the following we will refer to all these types
of showers as ``downward-going" $\nu$-induced showers.

In this review we restrict ourselves to downward-going $\nu$-induced showers with 
zenith angles $75^\circ \leq \theta \leq 90^\circ$.

\end{enumerate}

\begin{figure*}[htb]
\begin{center}
\noindent
\includegraphics [width=0.95\textwidth]{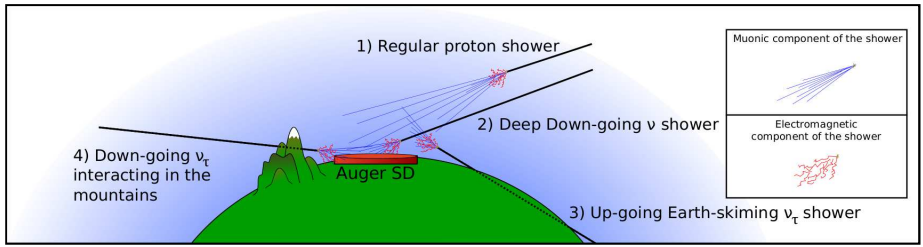}
\end{center}
\caption{
Pictorial representation of the different types of inclined showers that can be detected at 
the surface detector array of the \pao. 
(1) An inclined shower induced by a proton interacting high in the atmosphere whose 
electromagnetic component is absorbed and only the muons reach the detector.
Inclined showers presenting significant electromagnetic component at the detector level: 
(2) a deep downward-going $\nu$-induced shower; 
(3) an Earth-skimming $\nu_{\tau}$ interacting in the Earth crust and 
producing an upward-going $\tau$ lepton that decays in flight and induces a shower in the  
atmosphere; and (4) a $\nu_{\tau}$ interacting in the mountains, producing
a downward-going $\tau$ lepton that decays close to the detector and initiates a shower.
}
\label{fig:nu_sketch}
\end{figure*}

In Fig.~\ref{fig:nu_sketch} we show a pictorial representation of the 
different types of inclined showers that can be detected.

\section{Simulation of neutrino showers}
\label{sec:sims}

\begin{figure*}[htb]
\begin{center}
\noindent
\includegraphics [width=0.95\textwidth]{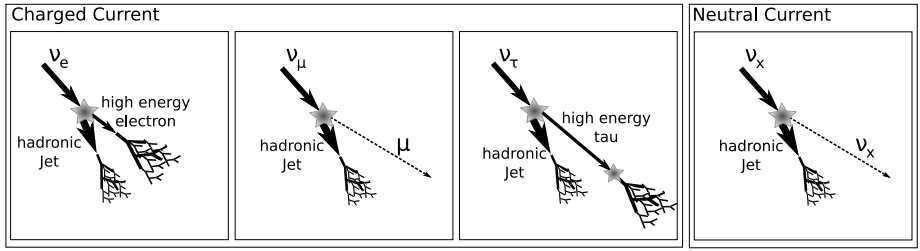}
\end{center}
\caption{Sketch of the different types of showers induced by UHE neutrinos.
All the channels depicted contribute to the neutrino event rate due to 
downward-going $\nu$ induced showers. 
}
\label{fig:nu_channels}
\end{figure*}

Monte Carlo simulations of neutrino-induced showers
are crucial to establishing identification criteria and computing
the acceptance of the SD to UHE$\nu$s. The whole simulation
chain is divided in three stages:

\begin{enumerate}

\item High energy processes:

\begin{enumerate}

\item The $\nu$-nucleon interaction in the atmosphere for downward-going neutrinos
is simulated with HERWIG \cite{HERWIG}.

The output of HERWIG includes the types, energies and momenta of the secondary particles
produced for both charged (CC) and neutral current (NC)
neutrino interactions (see Fig. \ref{fig:nu_channels} for a pictorial 
summary of all the channels considered in this work).

\item In the case of  $\nu_{\tau}$ CC interactions, the $\tau$ lepton propagation
in the Earth and/or in the atmosphere is simulated with a dedicated, fast and flexible
code which allows us to easily study the influence on the outgoing $\tau$ lepton flux
of different  $\nu_{\tau}$ interaction cross sections, $\tau$ energy loss models, etc.
The simulation of the decay of the $\tau$ (when necessary) is performed with
the TAUOLA package \cite{TAUOLA}.

\end{enumerate}

\item Shower development in the atmosphere:

The AIRES Monte Carlo code \cite{Aires} is used to propagate the particles produced in a high energy $\nu$ interaction,
or in the decay of a $\tau$ lepton. The types, energy-momenta and times of the particles 
reaching the SD level are obtained.  

\item Surface detector array simulation:

This is performed with the \Offline software \cite{Offline}.
Firstly particles reaching a surface detector station
are injected into the station, and with the aid of GEANT4 \cite{GEANT4} 
the amount of Cherenkov light produced in water is calculated. 
Then the FADC traces of the PMT signals are obtained, and 
the total signal due to the particles entering the station, 
as well as several quantities characterizing the FADC trace 
which will be relevant for neutrino identification are computed (see below). 
Also both the local trigger condition (T2 - either
Threshold or ToT), and the global trigger condition (T3) are applied 
to the simulated events in the same way as for collected data.

\end{enumerate}

The phase space of the simulations -- namely, neutrino energy,
zenith angle of incidence, interaction depth in the atmosphere for downward-going
neutrinos, and altitude of the $\tau$ decay in the case of Earth-skimming
\nutau -- spans a sufficiently wide range of numerical values as to guarantee that at the 
edges of the phase space none of the simulated showers
fulfills the global trigger conditions. This is taken as a clear indication that
a complete sample of showers has been produced without
introducing any bias and therefore that the Monte Carlo sample correctly represents
the characteristic of showers that could trigger the SD of the \pao.
For the Earth-skimming channel, showers were simulated at zenith angles between $90.1^\circ$ and $95.9^\circ$
and at an altitude of the decay point above the \pao$~$ up to 2500 m.
In the case of downward-going neutrinos, simulations were performed at zenith
angles in the range $75^\circ$--$89^\circ$.

\section{Identifying neutrino-induced showers}
\label{sec:nu_selection}

As stated above, the selection of potential neutrino-induced showers (neutrino candidates) 
is based on two steps:

\begin{enumerate}

\item Firstly, we select among the data collected at the SD of the \pao$~$ those events 
that arrive in inclined directions with respect to the vertical.

\item Secondly, we select among the inclined events those with FADC traces that are spread in time,
indicative of the presence of an inclined shower in the early stage of development,
a clear signature of a deeply interacting neutrino triggering the SD.

\end{enumerate}

Although the two steps above are the same for all the neutrino-induced showers searched
for at the \pao, due to the different nature of Earth-skimming and downward-going 
neutrino induced showers, the criteria and selection cuts that are applied to data are slightly different. 

\subsection{Selection of inclined events}

First of all, events occurring during periods of data acquisition instabilities
\cite{Auger_trigger} are excluded.

For the remaining events the FADC traces of the triggered stations are first
``cleaned'' to remove accidental signals induced (mainly) by atmospheric muons
arriving closely before or after the shower front -- produced in showers
different than the triggering one and which are below the energy threshold
of the Pierre Auger Observatory.
The trace cleaning procedure is detailed in \cite{nu_tau_long}.
After that, the start times of the signals in all stations included in the global trigger
are requested to be compatible with a plane shower front moving at roughly the speed of light.
This compatibility is realized through upper bounds on both, the largest residual
and the mean quadratic residual from the planar fit.
If the condition is not fulfilled, fits are attempted removing one station;
for this operation, the stations are sorted by increasing {\it quality}
(based on the integrated amplitude and the duration of the signal),
and the procedure is stopped as soon as a satisfactory solution is found.
If none is found, trials are made removing two stations, and so on.
The event is accepted if at least three (four) stations in the 
Earth-skimming (downward-going) case belong to the configuration.

\begin{figure*}[htb]
\begin{center}
\noindent
\includegraphics [width=0.95\textwidth]{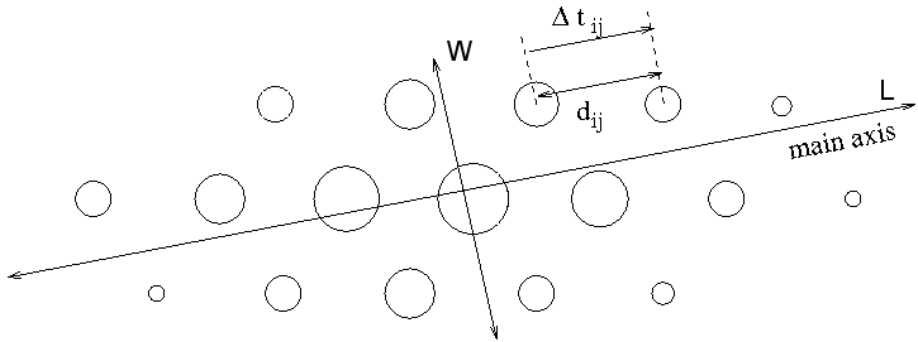}
\end{center}
\caption{Schematic view of the footprint of a shower triggering the
surface detector array of the \pao. The shower triggers the array from 
the left to the right of the figure, along the ``main axis". 
The circles represent the position of the stations, 
with their sizes being proportional to the collected signal in the PMTs. 
See text for more details.
}
\label{fig:footprint}
\end{figure*}

The second step in both channels is the selection of inclined showers. 
From the pattern (footprint) of stations at ground
(see Fig.~\ref{fig:footprint}) we can extract a length $L$ along the arrival direction
of the event (i.e., the main axis of the event) and a width $W$ perpendicular to it
characterizing the shape of the footprint
(see \cite{nu_tau_long} for complete details). The ratio $L/W$ depends on zenith
angle. Vertical events have $L/W\sim 1$ and this ratio increases
gradually as the zenith angle increases.
Very inclined events typically have elongated patterns on the
ground along the direction of arrival, and hence  
large values of $L/W$. A cut in $L/W$ is therefore a good selector of inclined events.
The exact value of this cut is different for downward-going 
and Earth-skimming events, and was determined through Monte Carlo
simulations of $\nu$-induced showers performed at different
zenith angles. For downward-going events with $\theta>75^\circ$ the 
requirement is $L/W>3$, while for Earth-skimming it is more restrictive
$L/W>5$ since only quasi-horizontal showers 
with largely elongated footprints can trigger the array
\footnote{
The axis of Earth-skimming showers travelling in
the upward direction does not intersect ground,
contrary to the downward-going showers case. 
For this reason, we exploit the properties of the
footprint generated by the shower particles that
deviate laterally from the shower axis and trigger
the water Cherenkov stations.}
(see Fig. 3 in Ref.\cite{nu_tau_long}). 

Another indication of inclined events is given by the apparent speed
$V$ of the trigger from a station $i$ to a station $j$, averaged over
all pairs $(i,j)$ of stations in the event. This observable denoted as \vavrg$~$ is obtained in
a straightforward manner from the distance
between the stations after projection along the ``main axis" of the footprint
at ground ($d_{ij}$) as depicted in Fig.~\ref{fig:footprint}, and from the difference in 
trigger times of the stations ($\Delta t_{ij}$). 
Vertical showers have apparent average speeds exceeding the speed of light since 
all triggers occur at roughly the same time, while in very inclined events 
\vavrg $~$is concentrated around the speed of light. Moreover its Root-Mean-Square 
(RMS($V$)) is small.
For downward-going (Earth-skimming) events \vavrg ~is required to be below 0.313 ${\rm m~ns^{-1}}$
(\vavrg $\in [0.29, 0.31]~{\rm m~ns^{-1}}$)
and RMS($V$)/\vavrg$<0.08$~(RMS($V$)$<0.08~{\rm m~ns^{-1}}$).
The values of these selection requirements are based on comparisons between data and Monte Carlo simulations.
Also, and only for downward-going events, a further quality cut is applied consisting on a 
simple reconstruction of the zenith angle $\theta_{\rm rec}$ and the requirement that  
$\theta_{\rm rec}>75^\circ$ (see \cite{nu_down} for full details).

In the top part of Table~\ref{tab:cuts} the cuts applied to the observables 
used to select inclined events are summarized. 

\begin{table}[hbt]
\begin{center}
\renewcommand{\arraystretch}{1.2}
\begin{tabular}{|c|c|c|}
\hline
          & Earth-skimming                                   & Downward-going \\
\hline\hline
          & Number of Stations $\geq$ 3                       & Number of Stations $\geq$ 4 \\
\hline
          & -                                                 & $\theta_{\rm rec}>$ 75$^{\circ}$ \\
Inclined  & $L/W > 5$                                         & $L/W > 3$ \\
Showers   & $0.29~{\rm m~ns^{-1}}<$\vavrg$<0.31~{\rm m~ns^{-1}}$ & \vavrg$~<~0.313~{\rm m~ns^{-1}}$ \\
          & RMS($V$)$~<~0.08~{\rm m~ns^{-1}}$                 & RMS($V$)/\vavrg$<0.08$ \\
\hline \hline
Young     & At least $60\%$ of stations with   & Fisher discriminant $\cal F$ based\\
Showers   & ToT trigger \& AoP $>$ 1.4        & on Area-over-Peak (AoP) \\
\hline
\end{tabular}
\end{center}
\caption{
Observables and numerical values of cuts applied to select {\it inclined} 
and {\it young} showers for Earth-skimming and downward-going neutrinos. See text for explanation.}
\label{tab:cuts}
\end{table}

\subsection{Selection of young showers} 
\label{sec:selection}

Once inclined showers are selected the next step is to identify young
showers among the data collected at the SD of the \pao. 

To optimize the numerical values of the cuts and tune 
the algorithms needed to separate neutrino-induced showers
from the much larger background of hadronic showers, we divided
the whole data sample 
into two parts (excluding periods of array instability). 
A fraction of the data (training period) is dedicated to define the selection algorithm.
These data are assumed to be overwhelmingly constituted of background showers.
The applied procedure is conservative because the
presence of neutrinos would result in a more severe definition of the selection criteria.
The remaining fraction is not used until the selection procedure
is established, and then it is ``unblinded" to search for neutrino candidates. 
In Table~\ref{tab:data_samples} we indicate the periods used 
for training and ``blind" search. 
The blind search period for the Earth-skimming
(downward-going) analysis corresponds to an equivalent of $\sim 3.5$ yr ($\sim 2$ yr) 
of a full surface detector array consisting of 1600 stations working continuously 
without interruptions. 

It is worth remarking that data  
instead of Monte Carlo simulations of hadronic showers are used to optimize
the identification cuts. The first reason for this is that,  
the composition of the primary UHECR flux -- a necessary input 
in the simulations -- is not accurately known.
Also, the detector simulation may not account for all possible 
detector defects and/or fluctuations that may induce events that 
constitute a background to UHE neutrinos, while they are
accounted for in collected data,
including those which are not well known, or even not yet diagnosed. 

This is the general strategy followed in the search for Earth-skimming \nutau and 
downward-going $\nu$-induced showers. However the two searches differ in several aspects
that we detail in the following sections.

\begin{table}
\begin{center}
\renewcommand{\arraystretch}{1.2}
\begin{tabular}{|l|c|c|}
\cline{2-3} 
\multicolumn{1}{c|}{}        & Earth-skimming       & Downward-going           \\
\hline 
Training period              & 1 Nov 04 - 31 Dec 04 & 1 Jan 04 - 31 Oct 07 \\
Blind search period          & 1 Jan 04 - 31 May 10 & 1 Nov 07 - 31 May 10 \\
\hline 
Equivalent full Auger        &                           &                      \\
blind search period          &  \raisebox{1.5ex}[0cm][0cm]{3.5 yr}  & \raisebox{1.5ex}[0cm][0cm]{2.0 yr} \\
\hline 
$\nu$ candidates & 0 & 0 \\
\hline 
Diffuse limit 90$\%$ C.L.         &                  &   \\
(${\rm GeV~cm^{-2}~s^{-1}~sr^{-1}}$) & 
\raisebox{1.5ex}[0cm][0cm]{$k~<~3.2\times 10^{-8}$}   &
\raisebox{1.5ex}[0cm][0cm]{$k~<~1.7\times 10^{-7}$}       \\
\hline 
Energy range (EeV)                & $\sim ~0.16 - 20.0$   &  $\sim ~0.1 - 100.0$  \\
\hline 
\end{tabular}
\end{center}
\caption{Training and blind search periods for the search for Earth-skimming and 
downward-going neutrino candidates. In the $3^{\rm rd}$ row we indicate the equivalent 
period of time of a full surface detector array.
In the $4^{\rm th}$ row we give the number of candidates found in the search period
after unblindly applying the cuts selecting {\it inclined} and {\it young} showers
(see Table~\ref{tab:cuts}).
In the $5^{\rm th}$ row we give the numerical value of 
the $90\%$ C.L. limit to the normalization $k$ of a diffuse flux of UHE neutrinos assumed 
to behave with energy as $dN/dE = k~E^{-2}$. Systematic uncertainties
are included in the value of the limit (see Sec. \ref{sec:sys} for details).
In the last row we indicate the energy range
where the limits apply, typically the energy interval where $90\%$ of the events are expected. 
}
\label{tab:data_samples}
\end{table}

\subsubsection{Earth-skimming analysis} 

In the Earth-skimming analysis we identify young showers by placing a cut
on the fraction of stations in the event that fulfill two conditions:
(1) the station passes the ToT local trigger condition, and
(2) the ratio of the integrated signal over the peak height
-- the so-called Area-over-Peak (AoP), a variable that carries information
on the time spread of the signal --
is greater than 1.4. By convention, both the ``area'' and the
``peak'' values are normalized to 1 in signals induced by isolated muons.

The aim of both conditions is to identify broad signals
in time such as those induced by showers developing close to the array.
In particular, with the second condition we reject background signals induced by
inclined hadronic showers, in which the muons and their electromagnetic products
are concentrated within a short time interval, exhibiting AoP values close to the one
measured in signals of isolated muons.

In order to reject inclined hadronic events, at least $60\%$ of the 
triggered stations in the event are required to fulfill the two conditions above 
(Table~\ref{tab:cuts}). The selection conditions were optimized using data 
collected during the training period indicated in Table~\ref{tab:data_samples}.
It is important to remark that this is the same selection procedure
and training period as in previous publications \cite{PRL_nu_tau, nu_tau_long},
which is applied in this work to a larger data set.
The final choice of the actual values of the neutrino selection cuts was 
done by requiring zero background events in the training data sample.
When the Earth-skimming cuts in Table~\ref{tab:cuts} are applied blindly to the data 
collected during the search period, no events survived. 

\subsubsection{Downward-going analysis} 

\begin{figure*}[htb]
\begin{center}
\noindent
\includegraphics [width=0.47\textwidth]{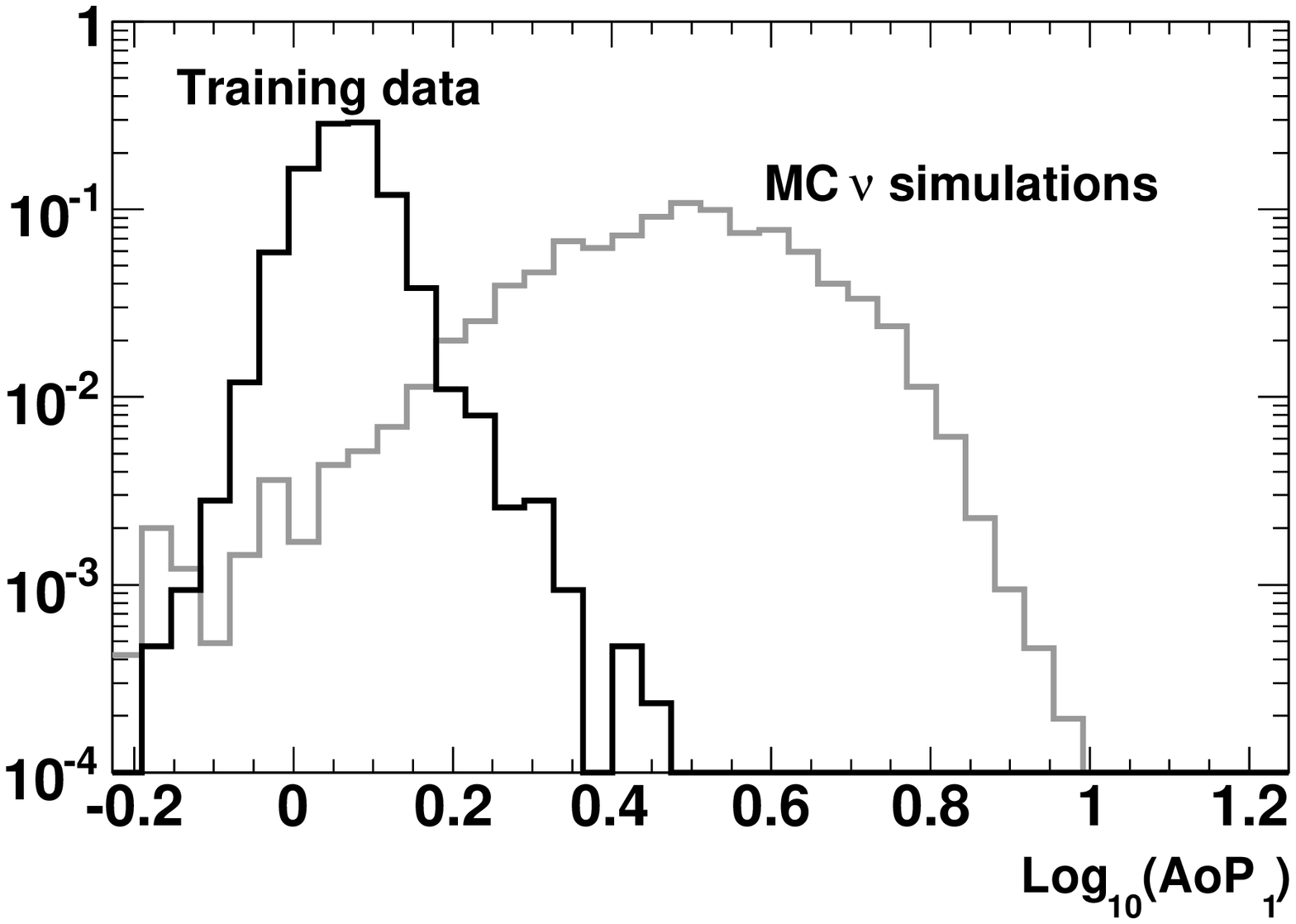}
\includegraphics [width=0.47\textwidth]{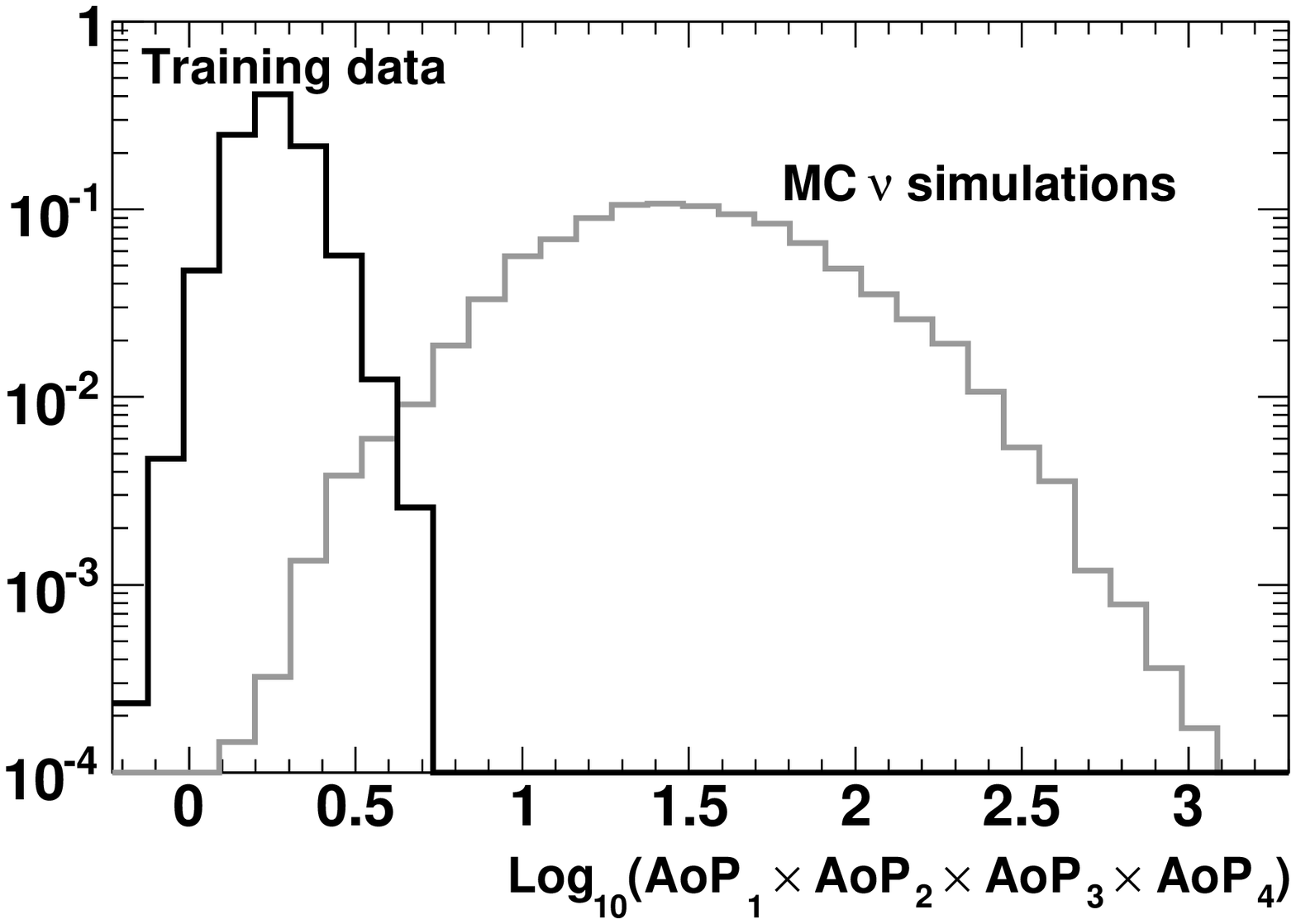}
\end{center}
\caption{Distributions of the Area-over-Peak (AoP, see text) of the earliest station (left) 
and of the product of the AoP of the first four stations in the event (right).
In each panel we show the distribution of the corresponding variable in background events 
(i.e. data events in the training sample as indicated in Table~\ref{tab:data_samples}), 
and in simulated electron neutrino charged-current events. 
These are two of the ten variables depending on the AoP used in constructing the multi-variate Fisher discriminant
linear polynomial to optimize the separation between background and neutrino-induced showers.
See text for more details on the remaining 8 variables.
}
\label{fig:AoP}
\end{figure*}

In the search for downward-going events, the discrimination power is optimized 
with the aid of a multi-variate technique known as the Fisher discriminant method \cite{Fisher}. 
The method consists on constructing a linear combination of observables 
denoted as $\cal F$ which optimizes the separation between
two samples of events, in our case background hadronic inclined showers occuring during 
the downward-going training period (see Table~\ref{tab:data_samples}), and Monte Carlo
simulated $\nu$-induced showers. The method requires as input a set of variables 
which can discriminate between the two samples. For that purpose 
we use variables depending on the Area-over-Peak (AoP) -- as defined above -- of the FADC traces. 
In the first few stations hit by a deep inclined shower, 
the typical AoP values range between 3 and 5 (left-hand panel of Fig.~\ref{fig:AoP}).

After training the Fisher method, a good discrimination
is found when the following ten variables are used to construct
the linear Fisher discriminant variable $\cal F$: the
AoP of the four stations that trigger first (early stations) in each event,
their squares, the product of the four AoPs, and a global parameter that
measures the asymmetry between the average AoP of the early
stations and those triggering last (late stations) of the event. 

The product of the AoP of the earliest four stations in the event aims at
minimizing the relative weight of an accidentally large
AoP produced, for instance, by a single muon which
does not belong to the shower front arriving at a station
before or after the shower itself. This variable is also a very
good discriminator as shown in the right-hand panel of Fig. \ref{fig:AoP}. 
We have also checked in Monte Carlo simulations that neutrino-induced
events typically have an asymmetry parameter larger than
proton or nucleus-induced showers. 

\begin{figure*}[htb]
\begin{center}
\noindent
\includegraphics [width=\textwidth]{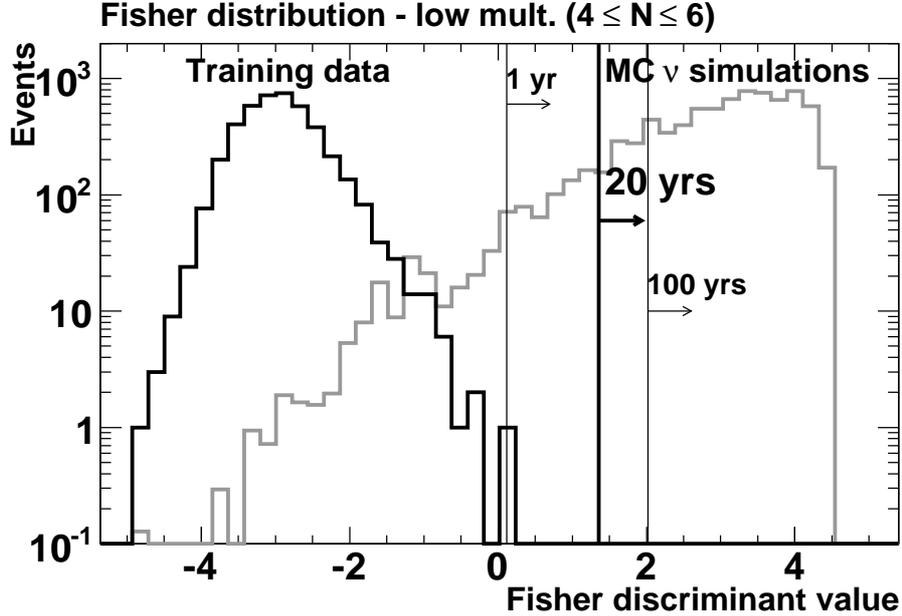}
\end{center}
\caption{
Distribution of the value of the Fisher polynomial ($\cal F$, see text for details) for events with 
number of selected stations $4 \leq N \leq 6$. Data in the training period 
(see Table~\ref{tab:data_samples}) describe the nucleonic background, 
while Monte Carlo simulated downward-going neutrinos correspond to the signal. 
The vertical lines indicate $\cal F_{\rm cut}$ needed to expect 
1 event in the labeled periods of time (full SD array).  
}
\label{fig:fisher}
\end{figure*}

As the shower front is broader at larger distance from the
core for both young and old showers, the discrimination is
better when splitting the samples according to
the number of selected stations $N$. A Fisher discriminant
polynomial was obtained separately for $4 \leq N \leq 6$, $7 \leq N \leq 11$, and
$N \geq 12$. An excellent separation is achieved for events in each of the three
sub-samples. The individual AoPs of the first four tanks have the largest weights in the 
Fisher polynomials.  
In Fig.~\ref{fig:fisher} we show as an example the distribution of $\cal F$ in the 
sub-sample with the smallest number of selected stations
(the distributions corresponding to the three sub-samples
can be found in Fig. 7 of \cite{nu_down}).

Once the Fisher discriminant $\cal F$ is defined, the next
step is to define a numerical value of $\cal F$, denoted
as ${\cal F}_{\rm cut}$, that separates neutrino candidates
from regular hadronic showers. 
One of the advantages of the Fisher discriminant method 
is that it allows us to estimate the expected rate of 
background events, and hence to tune the value of ${\cal F}_{\rm cut}$ 
so that the background is kept at a very low value. This is 
important given the fact that the 
expected rate of detected neutrino events will be small.
Data in the training period indicated in Table~\ref{tab:data_samples} 
was exploited to produce a reasonable prediction of the background
(see \cite{nu_down} for full details).
In practice, we fix $\cal F_{\rm cut}$ so that the estimated number
of background events is $1$ in 20 yr of data taking by a full Auger SD.  
With this cut, and for our search sample we have an estimated background
of 0.1 events for each multiplicity class that add up to a
total of 0.3 events with a statistical uncertainty of $30\%$. 
It is important to remark that this estimate relies on the a priori 
hypothesis that the background has an exponential distribution in $\cal F$.
Given the fact that we do not have a solid estimation of the actual background,
a conservative approach was taken assuming the background is zero, 
in other words, the estimated 0.3 background events were not used 
to improve our upper limit on the flux \cite{Feldman-Cousins} (see
Sec. \ref{sec:limits} below).

As exemplified in Fig.~\ref{fig:fisher} for the low multiplicity events, the identification cuts reject
only $\sim 10\%$ of the simulated neutrino events, and those are mainly neutrinos
interacting far from the ground that, being similar to nucleonic-induced showers,
are not expected to be identified.

Finally, when the downward-going cuts in Table~\ref{tab:cuts} are applied to the data 
collected during the search period, no neutrino candidates
appeared (see Table~\ref{tab:data_samples}).

\section{Exposure to UHE neutrinos}
\label{sec:exposure}

\subsection{Neutrino identification efficiencies}

With the criteria to select neutrino-induced showers indicated 
in Table~\ref{tab:cuts}, we obtain a relatively large identification 
efficiency both for Earth-skimming \nutau and downward-going $\nu$-induced showers. 
The efficiency has been computed with Monte Carlo simulations as the 
fraction of simulated events identified as neutrinos. 

\begin{figure*}[htb]
\begin{center}
\noindent
\includegraphics [width=\textwidth]{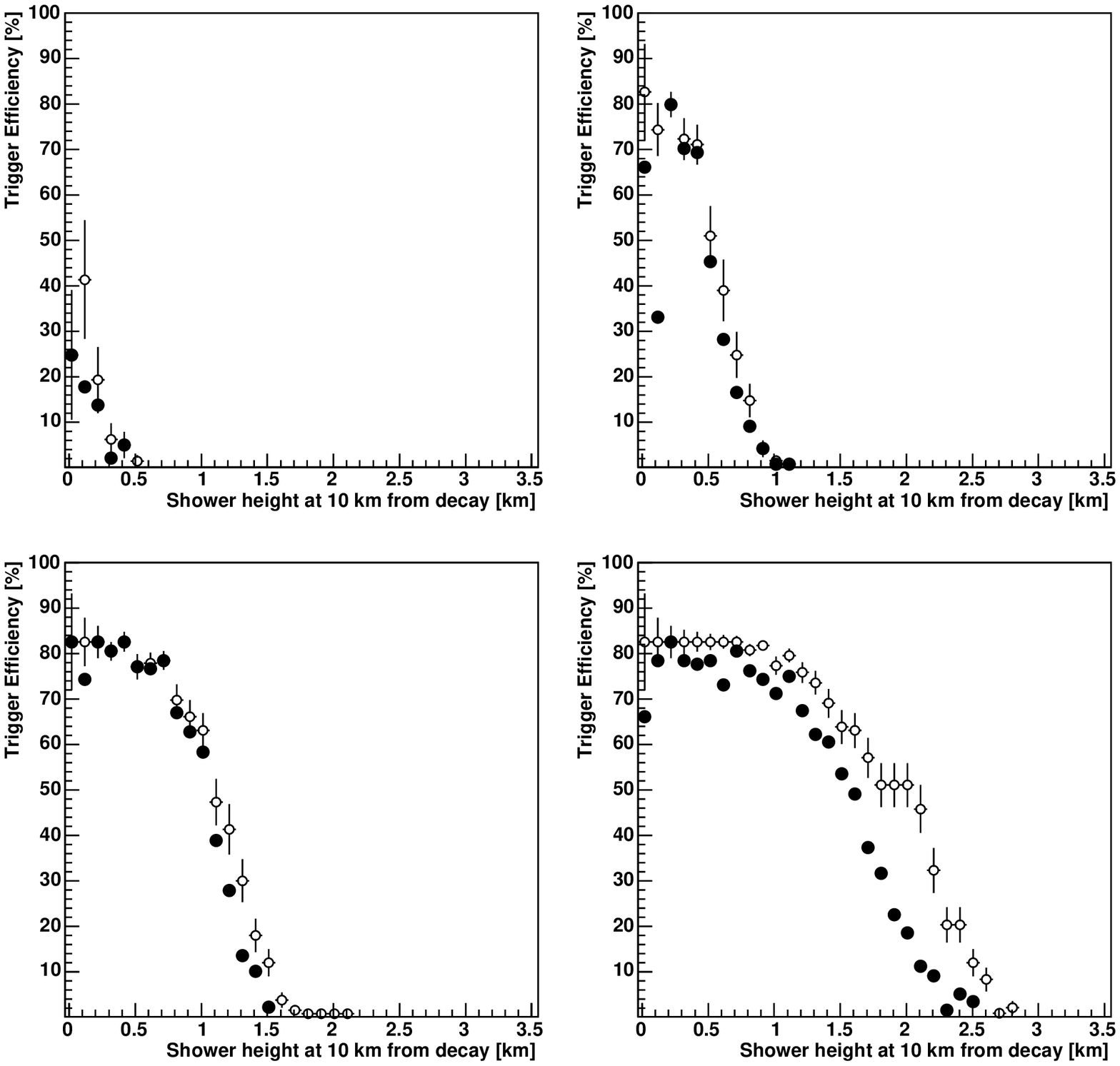}
\end{center}
\caption{
T3 trigger (open dots) and identification (closed dots, cuts as in Table~\ref{tab:cuts})
efficiency in the Earth-skimming analysis, as a function of the height above ground of the shower
at 10 km from the $\tau$ decay point $h_c$. The efficiency is shown for Monte Carlo
showers induced by $\tau$s with energy (clockwise from the top left panel) 
0.1, 1, 10 and 100 EeV. 
The efficiencies are calculated in a full SD array (see text for details).
}
\label{fig:effs_ES}
\end{figure*}

In the case of Earth-skimming \nutau induced showers, 
and a full Auger SD working without interruption,
the efficiencies  depend only on the energy of the 
emerging $\tau$ leptons ($E_\tau$) and on the altitude of the
``center of the shower" ($h_c$) above ground (averaged over the decay channels). This is conveniently 
defined as the altitude of the shower axis at a distance of 10 km away
from the $\tau$ decay point along the shower axis.
Showers induced by $\tau$ leptons with the 
same energy but with different zenith angles -- the range in $\theta$ being very narrow --
have approximately the same efficiency
as long as the corresponding altitudes of their shower maxima $h_c$ are the same.
The maximum efficiency that can be reached
is $82.6\%$, the $17.4\%$ remaining corresponds to the channel 
in which the $\tau$ decays into a $\mu$ which is unlikely to 
produce a detectable shower close to ground. 
In Fig.~\ref{fig:effs_ES} we show the 
trigger and identification efficiencies as a function of $h_c$ 
for different $\tau$ energies. As expected, the efficiency increases with 
$E_\tau$ and drops as the $\tau$ decays at increasing
altitude from ground.

\begin{figure*}[htb]
\begin{center}
\noindent
\includegraphics [width=\textwidth]{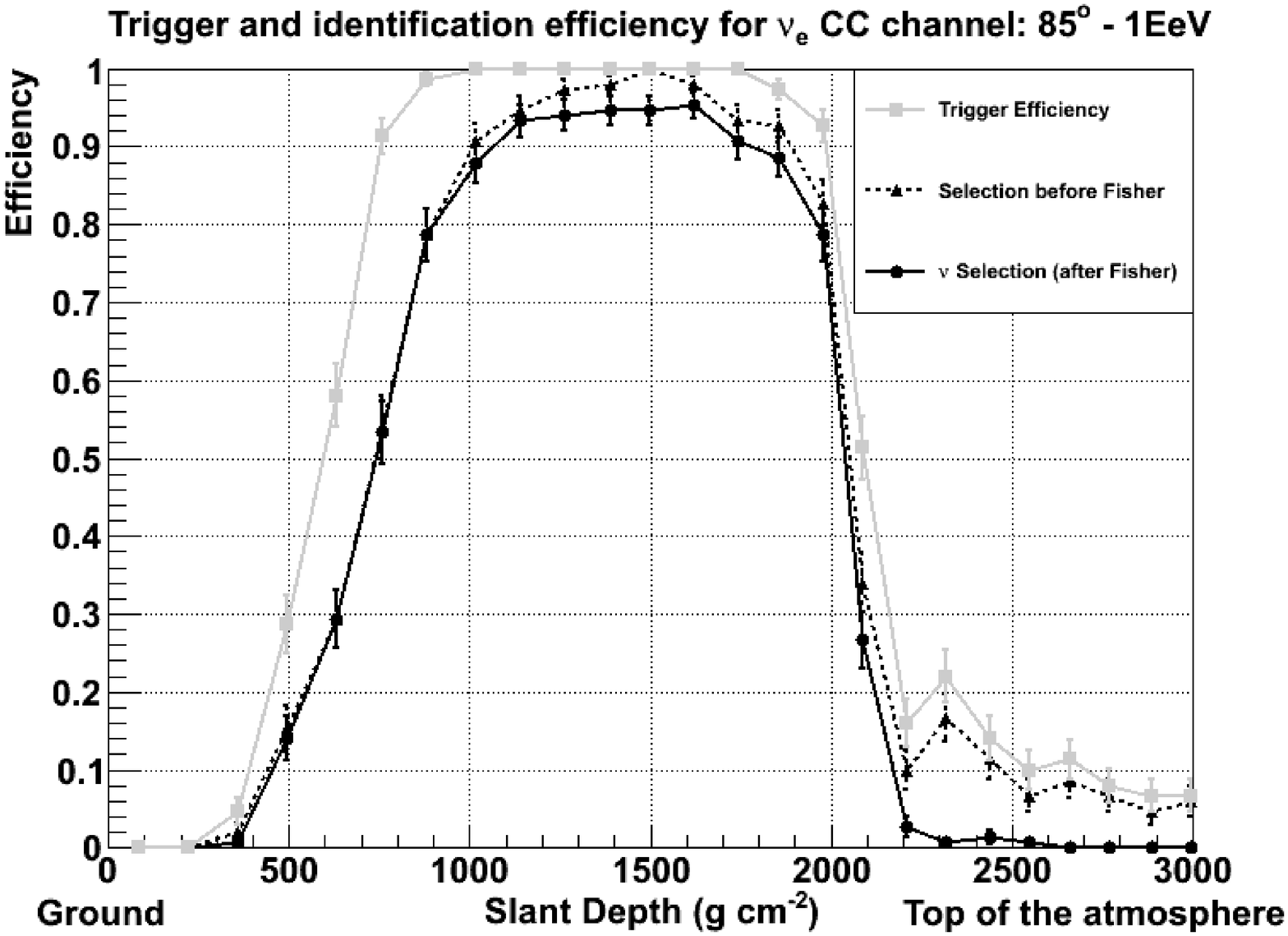}
\end{center}
\caption{
Fraction of electron neutrinos of energy 1 EeV and  $\theta=85^\circ$ 
triggering the array (solid grey line) and passing the downward-going analysis cuts in 
the second column of Table~\ref{tab:cuts}
(solid black line) as a function of the slant depth of the interaction above the ground.
The dashed line represents the fraction of events passing all cuts except for the 
cut on the Fisher discriminant $\cal F$ (see Sec. \ref{sec:selection}).
The efficiencies are calculated in a full SD array (see text for details).
}
\label{fig:effs_down}
\end{figure*}

In the case of downward-going neutrinos the identification efficiency
depends on neutrino flavour, type of interaction (CC or NC), neutrino 
energy ($E_\nu$), zenith angle ($\theta$), and distance ($D$) measured
from ground along the shower axis at which the neutrino is forced to interact
in the simulations. 
An example of the efficiency that can be achieved in a full SD array 
is shown in Fig.~\ref{fig:effs_down}.
The efficiency is different from zero between a minimal depth 
close to ground (a minimal amount of matter needed for the $\nu$-induced shower to reach a 
sufficient lateral expansion), and a maximal one (such that the 
electromagnetic component is almost extinguished at ground level and
hence the neutrino cannot be identified). 
The efficiency as well as the slice of atmosphere where it is different 
from zero, typically increase with neutrino energy, and depend on the 
neutrino flavour and interaction. As an extreme example, high energy \nutau interacting in 
the atmosphere through the CC channel 
can be identified regardless of the interaction depth in the atmosphere, 
as long as the energetic $\tau$ produced in the interaction decays and 
produces a shower close to ground.

\subsection{Exposure}

Ideally, for the calculation of the exposure of the SD of the Auger Observatory to
ultrahigh energy neutrinos, the simulated neutrino showers should be randomly
distributed over the actual configurations of the array, applying to the shower
at ground the trigger and neutrino identification conditions to obtain the active
(effective) area of the array at every second, and as a function of the parameters
of the neutrino-induced showers (neutrino energy, zenith angle, $h_c$,...). A sum over
time and integration in solid angle would then yield the {\it exposure} ($\cal E$) to UHE neutrinos
in both the Earth-skimming and downward-going neutrino analyses.
During the search periods considered for both Earth-skimming and downward-
going neutrino searches, the surface detector array of the Pierre Auger Observatory
was growing continuously. Since the number of working stations and their status
are monitored every second, we know with very good accuracy the SD configuration
at any instant as well as its evolution with time.
 
In practice, to avoid having to cope with an unaffordable number of configurations,
different strategies were devised to calculate in an accurate and less time-consuming
manner the effective area
of the SD array to Earth-skimming and downward-going $\nu$-induced showers. 

For downward-going neutrinos, the calculation of the exposure 
involves folding the SD array aperture with the $\nu$ interaction probability 
and the identification efficiency, and integrating in time. Changes
in the configuration of the array introduce a dependence
of the efficiency $\epsilon$ on the position of the core of the shower $\vec{r}=(x, y)$ in
the surface $S$ covered by the array, and on time $t$. 

Assuming a 1:1:1 flavour ratio (as expected due to the effects of neutrino oscillations during propagation from the sources), 
the total exposure can be written as \cite{nu_down}:
\begin{eqnarray}
\label{eq:exposure_down_diffuse}
{\cal E}^{\rm DG}(E_\nu)=\frac{2\pi}{m}\sum_i\left[ \sigma^i(E_\nu) \int\, \,dt\,d\theta\,dD\, \right.\,\,\,\,\,\,\,\,\,\,\,\,\,\,\nonumber\\ 
\left. \,\,\,\,\,\,\,\, \sin\theta\,\cos\theta\,\,A^{i}_{\rm eff}(\theta,D,E_\nu,t) \right]
\end{eqnarray}
where the sum runs over the three neutrino flavours and the CC and NC interactions,
with $\sigma^i$ the corresponding $\nu$-nucleon interaction cross-section \cite{Cooper-Sarkar} and $m$ the nucleon mass.
The integral is performed over the zenith angle $\theta$, the interaction depth $D$ of the neutrino (in units of ${\rm g~cm^{-2}}$), 
and the blind search period. $A^i_{\rm eff}$ is the effective area of the SD array given by:

\begin{eqnarray}
\label{eq:effective_area}
A^i_{\rm eff}(E_\nu,\theta,D,t)= 
\int \epsilon^{i}(\vec{r},\theta,D,E_\nu,t)~ dA
\end{eqnarray}
where the integral is performed over the core positions $\vec r$ of the showers.

For the Earth-skimming neutrinos the calculation of the exposure is described in \cite{nu_tau_long}.

\begin{figure*}[htb]
\begin{center}
\noindent
\includegraphics [width=\textwidth]{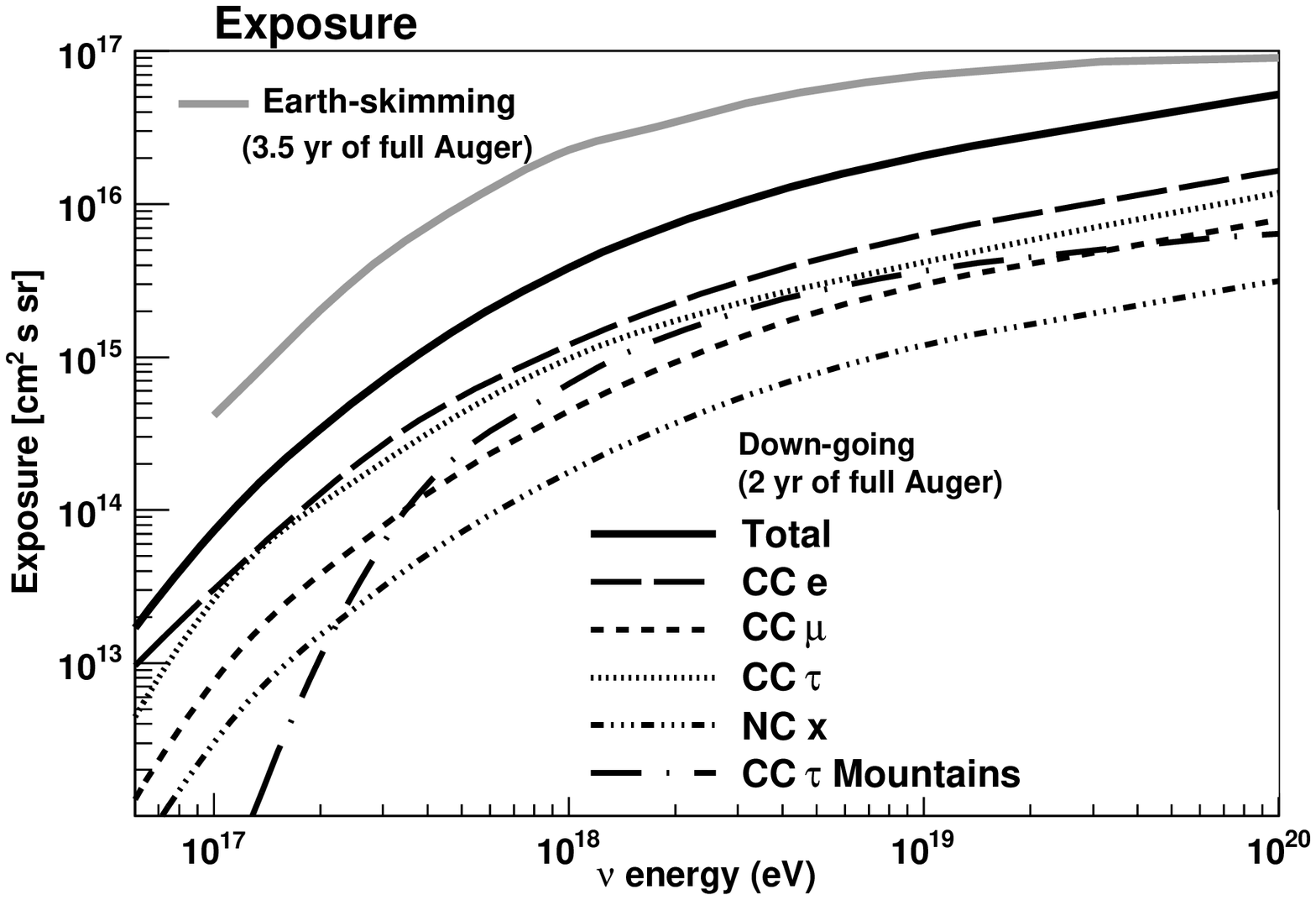}
\end{center}
\caption{
Exposure of the surface detector array of the Pierre
Auger Observatory on the data search periods
to Earth-skimming $\nu$-induced showers (equivalent to 3.5 yr of full Auger), 
and to downward-going $\nu$-induced showers (equivalent to 2 yr of full Auger).
}
\label{fig:exposure}
\end{figure*}

The exposures obtained for the search periods indicated in 
Table~\ref{tab:data_samples} are plotted in Fig.~\ref{fig:exposure}, 
where for the downward-going neutrino induced showers we also plot the contribution of the
different channels (Fig.~\ref{fig:nu_channels}) to the total exposure.
Among them we have included the possibility that downward-going $\nu_\tau$ interact
with the mountains surrounding the Observatory which provide a dense target
for neutrino interactions.  

The exposure to Earth-skimming neutrinos is higher than that to down\-ward\--going neutrinos by a factor between 
$\sim 2$ and $\sim 7$ depending on the neutrino energy,
partially due to the longer search period in the Earth-skimming 
analysis $\sim 3.5$ yr of full Auger, compared to $\sim 2.0$ yr in the case of the downward-going analysis. 
When normalized to the same search time, the Earth-skimming channel is still a factor $\sim 2.5-3$ 
more sensitive when integrated over the whole energy range, mainly due to the
larger density of the Earth's crust where \nutau interactions can occur, compared to the atmosphere. 
The larger number of neutrino flavours and interaction channels that can be identified in the downward-going analysis, 
as well as the broader angular range ($75^\circ<\theta<90^\circ$ compared to $90^\circ<\theta<95^\circ$), 
partly compensate the difference.  

\subsection{Systematic Uncertainties}
\label{sec:sys}

Several sources of systematic uncertainty have been carefully considered.
Some of them are directly related to the Monte Carlo simulation of the showers,
i.e., generator of the neutrino interaction either in the Earth or in the 
atmosphere, parton distribution function, air shower development, and
hadronic model. Others have to do with the limitations on the
theoretical models estimating, for instance, the interaction
cross-section or the $\tau$ energy loss at high energies.
Some of these sources play a dominant role on the Earth-skimming analysis, while
others do on the downward-going neutrino one. 

In both analyses the procedure to incorporate the systematic uncertainties
is the same. Different combinations of the various
sources of systematic uncertainty render different values of the exposure, 
and the final uncertainty is incorporated in the value of the limit itself 
through a semi-Bayesian extension \cite{Conrad} of the Feldman-Cousins approach \cite{Feldman-Cousins}.
In Table~\ref{tab:sys} we summarize the dominant sources of systematic uncertainty
and their impact on the exposure. In the Earth-skimming analysis the model of energy
loss for the $\tau$ is the dominant source of uncertainty, since it determines the 
energy of the emerging $\tau$s after propagation in the Earth; the impact of this on the downward-going analysis is 
much smaller since $\tau$ energy losses are only relevant for \nutau interacting
in the mountains, a channel that is estimated to contribute only $\sim15\%$ to the total exposure. 
The uncertainty on the shower simulation, that stems mainly from the different shower propagation codes
and hadronic interaction models that can be used to model the high energy collisions in the shower, 
contributes significantly in both cases. 
The presence of mountains around the Observatory -- which would increase the target for 
neutrino interactions in both cases -- is explicitly simulated and 
accounted for when obtaining the exposure of the SD to downward-going neutrino-induced showers,
and as a consequence does not contribute directly to the systematic uncertainties.  
However, it is not accounted for in the Earth-skimming limit shown 
in Table~\ref{tab:data_samples}. Instead, we take the topography around
the Observatory as a source of systematic uncertainty and we estimated that accounting for it 
would have increased the event rate by $\sim 18\%$ (Table~\ref{tab:sys}).

\begin{table}
\begin{center}
\renewcommand{\arraystretch}{1.2}
\begin{tabular}{|l|c|c|}
\hline 
Source of uncertainty            & Earth-skimming  & Downward-going \\
\hline 
Monte Carlo simulation of shower &  $+20\%,  -5\%$  & $+9\%, -33\%$ \\ 
$\nu$-nucleon cross-section      &  $ +5\%,  -9\%$  & $+7\%, -7\%$  \\ 
$\tau$ energy losses             &  $+25\%, -10\%$  & $+6\%, -6\%$  \\ 
Topography                       &  $+18\%,   0\%$  & --            \\ 
\hline 
\end{tabular}
\end{center}
\caption{Main sources of systematic uncertainty and their impact on the Earth-skimming \cite{nu_tau_long} 
and downward-going \cite{nu_down} exposures.}
\label{tab:sys}
\end{table}

\section{Results}
\label{sec:results}

We have searched for neutrino candidates over the search data periods and no events
fulfilling either the Earth-skimming or the downward-going selection cuts were found.
This allows us to put limits to the UHE diffuse neutrino flux.

\subsection{Limits to the diffuse flux of UHE neutrinos}
\label{sec:limits}

Under the assumption that the UHE neutrino flux $\Phi(E)$ behaves with neutrino energy as:

\begin{equation}
\Phi(E)=\frac{dN}{dE} = k~E^{-2} ~~~~~~ {\rm [GeV^{-1}~cm^{-2}~s^{-1}~sr^{-1}]} \; \; ,
\end{equation}

\par\noindent
the {\it integrated} limit on the value of $k$ is:

\begin{equation} 
k = \frac{N_{\rm up}}{\int_{E_{\rm min}}^{E_{\rm max}}~E^{-2}~{\cal E}(E)~dE}
\end{equation}

\par\noindent
where ${\cal E}$$(E)$ is the exposure.
The actual value of the upper limit on the signal events ($N_{\rm up}$)
depends on the number of observed and expected background events.
We recall here that, according to \cite{Feldman-Cousins},
$N_{\rm up}=2.44$ at 90\% C.L. for zero candidates and no expected background events.
When systematic uncertainties are included (Sec. \ref{sec:sys})
the value of $N_{\rm up}$ changes.

The final limits are reported in Table~\ref{tab:data_samples}
where we give the normalization $k$ obtained in the search 
periods (indicated in the same table) for the Earth-skimming and downward-going searches. 

\begin{figure*}[htb]
\begin{center}
\noindent
\includegraphics [width=\textwidth]{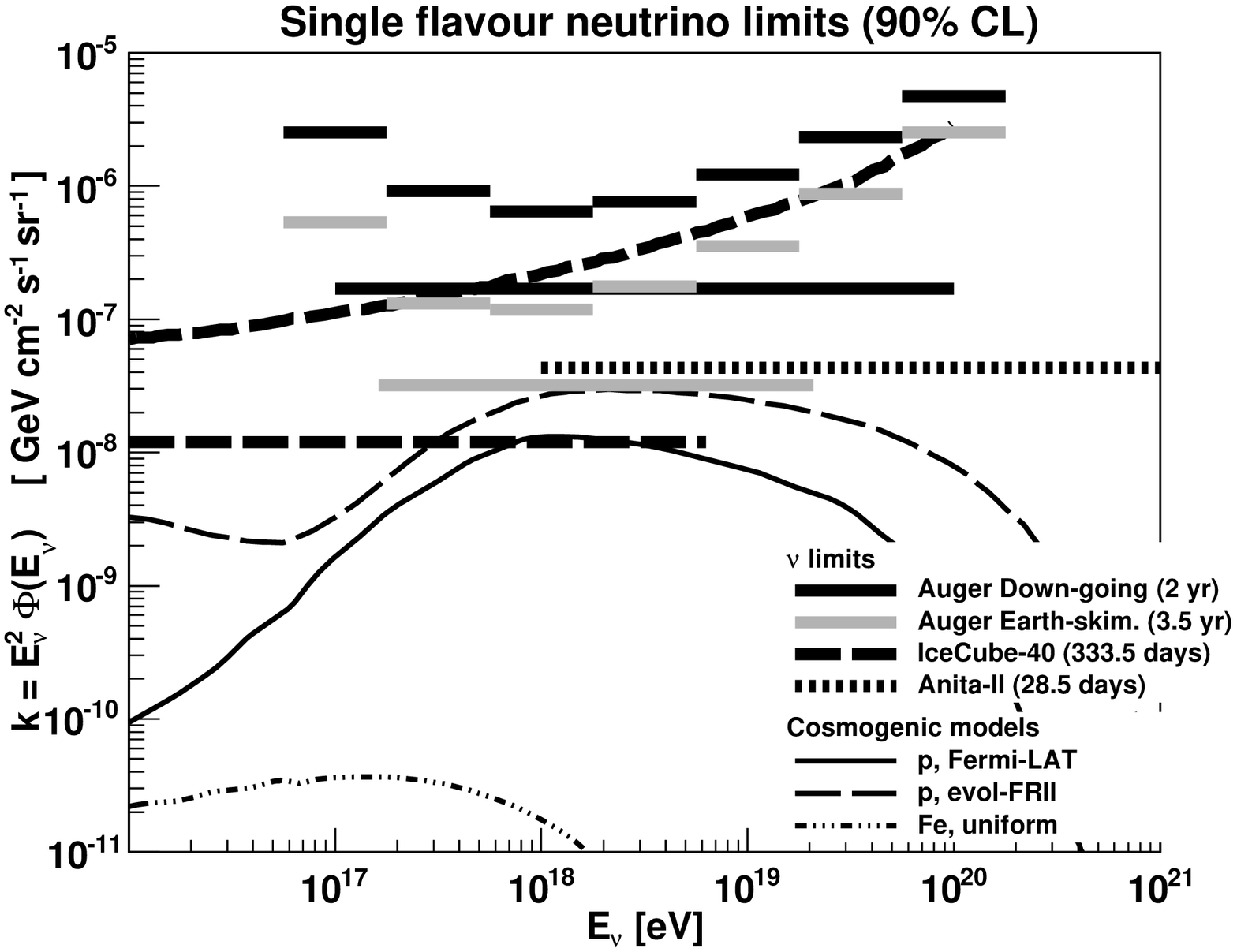}
\end{center}
\caption{
Thick lines: Differential and integrated upper limits (at $90\%$
C.L.) to the diffuse flux of UHE neutrinos (single flavour assuming 
equipartition) from the Pierre Auger Observatory 
for downward-going $\nu$ (equivalent search period = 2 yr of full Auger) 
and Earth-skimming \nutau (equivalent search period = 3.5 yr of full Auger). 
Limits from other experiments are also plotted \cite{IceCube,ANITA}. 
All limits have been scaled to single flavour. The IceCube differential limit
is scaled by a factor $1/2$ due to the different binning in energy with respect 
to the Auger differential limits. 
\newline
Thin lines: Expected fluxes for 
three theoretical models of cosmogenic neutrinos (scaled to single flavour when necessary).
``p, Fermi-LAT'' \cite{Ahlers_GZK} corresponds to the best fit to UHECR spectrum incorporating Fermi-LAT
bound assuming that the transition from Galactic to extragalactic CRs takes place at $10^{19}$ eV.
``p, evol-FRII'' \cite{Kotera_GZK} assumes the FRII strong source evolution with a pure proton composition, 
dip transition model and maximum energy of UHECRs at the sources $E_{p,{\rm max}} = 10^{21.5}$ eV. 
``Fe, uniform'' \cite{Kotera_GZK} represents an extreme model assuming an iron rich composition,
low $E_{p,{\rm max}}$, uniform evolution of the UHECR sources case.
}
\label{fig:limits}
\end{figure*}

In Fig.~\ref{fig:limits} we show the Earth-skimming and downward-going 
integrated neutrino flux which indicate the level of a diffuse neutrino flux assumed to behave 
with energy as $E^{-2}$,  
needed to detect $N_{\rm up}$ events with a Poisson probability of $\sim 90\%$ given the exposure 
accumulated during the 3.5 years for Earth-skimming neutrinos (2.0 years for downward-going) of equivalent 
time of a full SD. 

Another way of presenting the results is to display the upper limit in
differential form. In this procedure we assume that the diffuse neutrino 
flux behaves as $E^{-2}$ within energy
bins of 0.5 width on a decimal logarithmic scale, and is given by 
$2.44/(0.5 \log(10) \cdot E \cdot {\cal E}(E))$, assuming again no background. The differential
limit obtained in this way is shown in Fig.~\ref{fig:limits} for the Earth-skimming 
and downward-going cases.
We achieve most ($\sim 90\%$) of the sensitivity in the energy range $\sim~ 0.16-20$ EeV ($\sim~ 0.1-100$ EeV) 
for Earth-skimming (downward-going) neutrinos.
In Fig.~\ref{fig:limits} we also show several predictions of different theoretical models
of cosmogenic neutrino production \cite{Ahlers_GZK,Kotera_GZK}.
Predictions for cosmogenic neutrino fluxes depend on several unknown parameters including
the evolution with redshift of the sources and the injected UHECR composition.
Given the uncertainties in these parameters, and in particular the possible presence of 
heavy primaries in the UHECR spectrum \cite{Auger_Xmax}, 
we have plotted a range of models 
to illustrate the wide range of predictions available \cite{Kotera_GZK}.   

\subsection{Event rate predictions}

In Table~\ref{tab:rates} we give the expected number of events from a diffuse 
flux of cosmogenic neutrinos (produced in the interaction of cosmic ray protons with 
background radiation fields) \cite{Ahlers_GZK}, from a model of neutrino production
through the bottom-up mechanism in Active Galactic Nuclei (AGN) \cite{BBR}, 
and from a theoretical model \cite{Sigl} in which neutrinos 
are the product of the decay of super-heavy relic particles of the early stages 
of the Universe. 
Optimistic theoretical flux predictions for cosmogenic neutrinos are within reach of our
present sensitivity and some models of neutrinos produced in accelerating
sources are already being constrained.
Exotic models are severely disfavored. Note that
all such ‘‘top down’’ models are also tightly constrained by
the limits of the Pierre Auger Observatory on the photon
fraction in UHECR \cite{Auger_photon_limit}. 

\begin{table}
\begin{center}
\renewcommand{\arraystretch}{1.2}
\begin{tabular}{|l|c|c|}
\hline
     Model $\&$ reference            & Earth-skimming  & Downward-going \\
\hline 
Cosmogenic (Fermi) \cite{Ahlers_GZK} & $\sim ~0.6$     & $\sim ~0.1$    \\ 
AGN \cite{BBR}                       & $\sim ~5.1$     & $\sim ~0.8$    \\
Exotic (SH relics) \cite{Sigl}       & $\sim ~3.0$     & $\sim ~1.0$    \\
\hline 
\end{tabular}
\end{center}
\caption{ Number of expected events for several theoretical models of UHE neutrino 
production, given the exposure of the surface detector array
of the \pao$~$ to Earth-skimming and downward-going neutrinos (Table~\ref{tab:data_samples}).}
\label{tab:rates}
\end{table}

\section{Summary and Prospects}
\label{sec:conclusions}

In this paper we have reviewed the searches for astrophysical sources of
ultrahigh energy neutrinos at the Pierre Auger Observatory~\cite{PRL_nu_tau,nu_tau_long,nu_down}.

The neutrino detection technique is based on the observation 
of extensive air showers induced by downward-going neutrinos of all flavours
as they interact with the atmosphere, and by upward-going $\nu_{\tau}$'s
through the Earth-skimming mechanism.
These $\nu$-induced showers display characteristic features that allow us
their identification in the overwhelming background of regular UHE hadronic showers.
At ground level, high zenith angle neutrino events would have a significant electromagnetic component
leading to a broad time structure of detected signals in the surface detector array,
in contrast to nucleonic-induced showers.

We have shown that, using Monte Carlo simulations and training data samples,
identification criteria for UHE neutrinos can be defined and used to perform
a blind search on the remaining data sample.
The analysis of the collected data at the Pierre Auger Observatory
until 31 May 2010 reveals no candidate events for either downward-going
or Earth-skimming neutrinos. Based on this negative result,
stringent limits have been placed on the diffuse flux of UHE neutrinos.
Even though the Auger Observatory was designed to measure properties of UHECRs,
the limits reported in Table~\ref{tab:data_samples}
provide at present one of the most sensitive bounds
on neutrinos at EeV energies,
which is the most relevant energy to explore the predicted
fluxes of cosmogenic neutrinos.

There are several lines of work in progress inside the Auger Collaboration
related to the neutrino search which will be the subject of future reports.
Some of the efforts concentrate on the combination of the downward-going
and Earth-skimming channels into a single analysis. This will simplify
the search procedure and will obviously translate into an
improvement of the diffuse neutrino limit.
The extension of the downward-going neutrino search to lower zenith angles
($\theta<75\degree$) is also very promising. 
Exploring the sky down to $\theta \sim 60\degree$ implies a sizeable
increase on the exposure and hence on the limit in case no candidates are found.
The main drawback of decreasing $\theta$ is that the atmosphere
slant depth reduces and nucleonic-induced showers look ``younger'' when
arriving at ground, making their separation from $\nu$-induced showers
more challenging.
On the other hand, the sensitivity to neutrino detection could also be
extended to lower energies by reducing the separation between SD stations.
Monte Carlo studies indicate that using a configuration of stations similar to the
currently existing ``infill'' array ($\sim 60$ stations spaced by 750 m)
would lead to a significant increase of the neutrino detection probability
at lower energies (below 0.3 EeV) with respect to the standard SD array.
Nevertheless, due to the small size of the current infill array,
the exposure does not appear to be competitive.

Finally, it is worth mentioning that the sensitivity of the Pierre Auger Observatory to the detection of UHE$\nu$s
from potential astrophysical {\it point-like} sources is being evaluated. The absence of candidates in the searches
for diffuse neutrino fluxes
described in this report allows us to place limits on the neutrino fluxes coming from sources in the field of view of the
SD of the Auger Observatory. Preliminary results indicate that with the SD we are sensitive to a large fraction
of the sky spanning $\sim 100\degree$ in declination \cite{PSinprep}.

\section{Acknowledgements}
\label{sec:acknowledgements}

The successful installation, commissioning, and operation of the Pierre Auger Observatory
would not have been possible without the strong commitment and effort
from the technical and administrative staff in Malarg\"ue.

We are very grateful to the following agencies and organizations for financial support: 
Comisi\'on Nacional de Energ\'ia At\'omica, 
Fundaci\'on Antorchas,
Gobierno De La Provincia de Mendoza, 
Municipalidad de Malarg\"ue,
NDM Holdings and Valle Las Le\~nas, in gratitude for their continuing
cooperation over land access, Argentina; 
the Australian Research Council;
Conselho Nacional de Desenvolvimento Cient\'ifico e Tecnol\'ogico (CNPq),
Financiadora de Estudos e Projetos (FINEP),
Funda\c{c}\~ao de Amparo \`a Pesquisa do Estado de Rio de Janeiro (FAPERJ),
Funda\c{c}\~ao de Amparo \`a Pesquisa do Estado de S\~ao Paulo (FAPESP),
Minist\'erio de Ci\^{e}ncia e Tecnologia (MCT), Brazil;
AVCR AV0Z10100502 and AV0Z10100522, GAAV KJB100100904, MSMT-CR LA08016,
LG11044, MEB111003, MSM0021620859, LA08015 and TACR TA01010517, Czech Republic;
Centre de Calcul IN2P3/CNRS, 
Centre National de la Recherche Scientifique (CNRS),
Conseil R\'egional Ile-de-France,
D\'epartement  Physique Nucl\'eaire et Corpusculaire (PNC-IN2P3/CNRS),
D\'epartement Sciences de l'Univers (SDU-INSU/CNRS), France;
Bundesministerium f\"ur Bildung und Forschung (BMBF),
Deutsche Forschungsgemeinschaft (DFG),
Finanzministerium Baden-W\"urttemberg,
Helmholtz-Gemeinschaft Deutscher Forschungszentren (HGF),
Ministerium f\"ur Wissenschaft und Forschung, Nordrhein-Westfalen,
Ministerium f\"ur Wissenschaft, Forschung und Kunst, Baden-W\"urttemberg, Germany; 
Istituto Nazionale di Fisica Nucleare (INFN),
Ministero dell'Istruzione, dell'Universit\`a e della Ricerca (MIUR), Italy;
Consejo Nacional de Ciencia y Tecnolog\'ia (CONACYT), Mexico;
Ministerie van Onderwijs, Cultuur en Wetenschap,
Nederlandse Organisatie voor Wetenschappelijk Onderzoek (NWO),
Stichting voor Fundamenteel Onderzoek der Materie (FOM), Netherlands;
Ministry of Science and Higher Education,
Grant Nos. N N202 200239 and N N202 207238, Poland;
Funda\c{c}\~ao para a Ci\^{e}ncia e a Tecnologia, Portugal;
Ministry for Higher Education, Science, and Technology,
Slovenian Research Agency, Slovenia;
Comunidad de Madrid, 
Consejer\'ia de Educaci\'on de la Comunidad de Castilla La Mancha, 
FEDER funds, 
Ministerio de Ciencia e Innovaci\'on and Consolider-Ingenio 2010 (CPAN),
Xunta de Galicia, Spain;
Science and Technology Facilities Council, United Kingdom;
Department of Energy, Contract Nos. DE-AC02-07CH11359, DE-FR02-04ER41300,
National Science Foundation, Grant No. 0450696,
The Grainger Foundation USA; 
NAFOSTED, Vietnam;
ALFA-EC / HELEN and UNESCO.



\end{document}